\documentclass[a4paper]{jpconf}

\usepackage{graphicx}
\usepackage{subscript}
\usepackage{amsmath}
\usepackage{amssymb}
\usepackage{textcomp}

\newcommand{\NH}{$\mathcal{NH}$} 
\newcommand{\IH}{$\mathcal{IH}$} 

\begin{document}

\title{Assess the neutrino mass with micro and macro calorimeter approach}
\author{A~Giachero}

\address{INFN Milano-Bicocca, Piazza della Scienza 3, Milano I-20126 - Italy}
\ead{andrea.giachero@mib.infn.it}

\begin{abstract}
Thanks to oscillation experiments it is now an established fact that neutrinos are massive particles. Yet, the assessment of neutrinos absolute mass scale is still an outstanding challenge in particle physics and cosmology as oscillation experiments are sensitive only to the squared mass differences of the three neutrino mass eigenstates. The mass hierarchy is not the only missing piece in the puzzle. Theories of neutrino mass generation call into play Majorana neutrinos and there are experimental observations pointing toward the existence of sterile neutrinos in addition to the three weakly interacting ones. Three experimental approaches are currently pursued: an indirect neutrino mass determination via cosmological observables, the search for neutrinoless double $\beta$-decay, and a direct measurement based on the kinematics of single $\beta$ or electron capture decays. 

Bolometers and calorimeters are low temperature detectors used in many applications, such as astrophysics, fast spectroscopy and particle physics. In particular, sensitive calorimeters play an important role in the neutrino mass measurement and in the search for the neutrinoless double $\beta$-decay.
There has been great technical progress on low temperature detectors since they were proposed for neutrino physics experiments in 1984. This general detector paradigm can be implemented in devices as small as a micrometer for sub eV radiation or as large as 1\,kg for MeV scale particles. Today this technique offers the high energy resolution and scalability required for leading edges and competitive experiments addressing the still open questions in neutrino physics.

\end{abstract}
After 30 years of observations, oscillation experiments have firmly established neutrinos are massive fermions whose three weak interaction flavor eigenstates ($\nu_e$, $\nu_\mu$, $\nu_\tau$) are a quantum superposition of three mass eigenstates ($\nu_1$, $\nu_2$ , $\nu_2$). Thanks to the accurate characterization of the neutrino oscillations, observed for solar, atmospheric, accelerator and nuclear reactor neutrinos, the neutrino mixing matrix $U_{\ell i}$ is now precisely determined. Neutrino oscillations are sensitive only to the difference of the squared masses $\Delta m^2_{ij} = | m^2_i- m^2_j |$. Two mass square differences have been determined reasonably well: $\Delta m^2_{12}\simeq (7.37\pm 0.17)\cdot 10^{-5}\,\mbox{eV}^2$ and $|\Delta m_{23}|^2\simeq |\Delta m_{13}|^2 \simeq (2.50\pm 0.04)\cdot 10^{-3}\,\mbox{eV}^2$, but the sign of $\Delta m_{13}^2$ is however unknown~\cite{Capozzi2016}. This sign settles the neutrino mass hierarchy: if it is positive the hierarchy is normal (\NH) with $m_1\ll m_2\ll m_3$ on the contrary if it is negative the hierarchy is inverted (\IH), with $m_3\ll m_1\lesssim m_2$. If $m_1\lesssim m_2\lesssim m_3$ the pattern is quasi-degenerate. Past oscillation experiments were not sensitive to the neutrino mass hierarchy. Currently, optimistic efforts to determine the neutrino mass hierarchy are basically accelerator based long-baseline (i.e. NO$\nu$A~\cite{NOvA2016} and T2K~\cite{T2K2013}) and reactor (i.e. JUNO~\cite{JUNO2016}) experiments. 

As alternative to oscillation experiments three experimental methods can assess the neutrino masses and the neutrino mass scale: the analysis of Cosmic Microwave Background (CMB) temperature fluctuations~\cite{Planck2016}, the search for the neutrinoless double $\beta$-decay ($0\nu\beta\beta$), and the analysis of the single $\beta$ or electron capture (EC) decay. The first method can give important indirect information on the neutrino mass by using space-borne instruments to precisely measure the CMB thermal fluctuations. These bounds depend on the delicate interplay between CMB and galaxy power spectra. This approach depends critically on cosmological and astrophysical assumptions and requires therefore independent checks. The observable of these measurements is the sum of the neutrino masses: $\Sigma ={\scriptscriptstyle\sum_{i=1}^3} m_i$, where $m_i$ are the mass eigenvalues of the three neutrino mass eigenstates. The most stringent limit is $\Sigma < 0.12\,\mbox{eV}$\,@\,95\%\,C.L.~\cite{Palanque-Delabrouille2015}. The second method has been suggested since long time as a powerful tool to measure electron neutrino effective mass, although its very existence depends on the assumption that the neutrino is Majorana particle ($\overline{\nu}\equiv\nu$). The critical problems of this approach are the theoretical uncertainties in the calculations of the nuclear matrix elements (NME) and phase space factors (PSF), that enter the $0\nu\beta\beta$ lifetime expressions~\cite{DellOro2016}. The observable is the effective neutrino Majorana mass: $m_{\beta\beta}=|{\scriptscriptstyle\sum_{i=1}^3 m_i\eta_i |U_{ei}|^2}|$, where $\eta_i$ are the CP Majorana phases ($\eta_i = \pm 1$ for CP conservation) and  $U_{ei}$ are the elements of the electron sector of the neutrino mixing matrix. The most stringent limits depend on the isotope and up-to-dately are: $m_{\beta\beta}(\mbox{\textsuperscript{76}Ge})\leq(160\div 260)$~meV~\cite{GERDANeutrino2016}, $m_{\beta\beta}(\mbox{\textsuperscript{130}Te})\leq(260\div 760)$~meV~\cite{CUORE0,CUORE0Analysis}, $m_{\beta\beta}(\mbox{\textsuperscript{136}Xe})\leq(61\div 165)$~meV~\cite{KamLAND-Zen}. In the last method the neutrino mass can be directly assessed by studying the kinematics of low energy nuclear $\beta$ or EC decays, where the neutrino is not directly observed but the energy of the decay products is precisely measured~\cite{Drexlin2013}. In these experiments the observable is the electron neutrino mass: $m_{\beta}=\scriptscriptstyle\sqrt{\sum_{i=1}^3 m_i^2 |U_{ei}|^2}$. Experiments based on the classical spectrometer technique provided the most stringent limits of $m_{\beta}<2\mbox{eV}$\,@\,90\%\,C.L.~\cite{Troitsk2001,Mainz2004}.

The investigation of fundamental neutrino properties, like its mass and its nature, calls for the design of a new generation of experiments. Different detection techniques and isotopes are being actively promoted by experimental groups
across the world. Low temperature detection (LTD) has been suggested as high resolution soft X-ray detectors in 1984 by D. McCammon and collaborators~\cite{McCammon1984}. The same year E. Fiorini and T.O. Niinikoski proposed LTD for neutrino physics experiments~\cite{Fiorini1984}. After more than 20 years of developments, the low temperature techniques are now ready for challenging the most pressing open questions in neutrino physics. Today LTD are widely used for neutrinoless double $\beta$-decay searches, for calorimetric direct neutrino mass experiments, but they play also an important role for dark matter searches~\cite{SuperCDMSdet} and X- and $\gamma$-ray spectroscopy in general~\cite{Ullom2015}. 

\begin{figure}[!t]
  \begin{center}
    \includegraphics[clip=true,width=\textwidth]{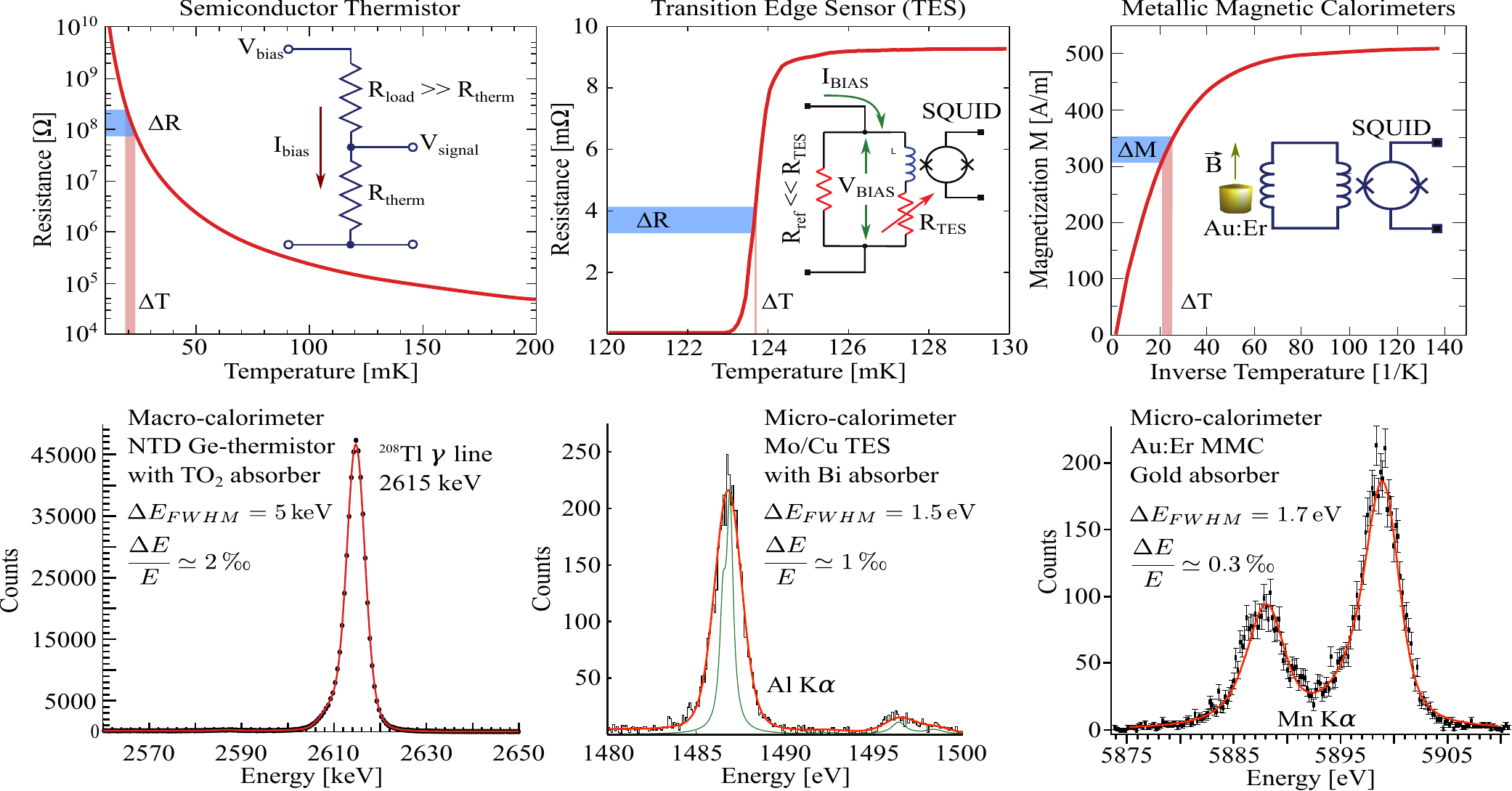}
  \end{center}
  \caption{\label{fig:sensors} Different sensors with temperature behaviour and energy resolutions. The TES spectrum (middle column) is courtesy of NIST. The thermistor (left column) and MMC (right column) spectra are edited from~\cite{CUORE0} and~\cite{Porst2014}, respectively.}
\end{figure}

\section{Low temperature micro- and macro-calorimeters}
In general, a low temperature calorimeter consists of: an \textit{energy absorber}, in which the energy deposited by a particle is converted into phonons, a \textit{temperature transducer} (sensor) that converts thermal excitations
into a readable electric signal variation, and a weak \textit{thermal link}. 
The temperature rise $\Delta T$ is related to the energy release $E$ and, denoting by $C$ the heat capacity of the absorber, can be written as $\Delta T = E/C$. The accumulated heat flows then to the heat bath through the thermal link and the absorber returns to the base temperature with a time constant $\tau = C/G$ where $G$ is the thermal conductance of the link: $\Delta T(t) = E/C\cdot e^{−t/\tau}$. The decay constant $\tau$ should be long enough to allow the development of a high resolution signal but short enough to avoid pile-up. In order to obtain a measurable temperature, the heat capacity and the base temperature of the absorber crystal must be very small. For these reasons low temperature detectors are usually operated at a temperature around 10\,mK and are made of superconductor or dielectric materials, so that, by the Debye law ($C\propto (T/\Theta_D)^3$), only the lattice heat capacity plays a role. The temperature variation measured by a low temperature calorimeter provides a precise estimation of the released energy, limited only by the so-called thermodynamic limit: $\Delta E_{rms}=\sqrt{k_BT^2C}$, where $T$ is the equilibrium operating temperature and $k_B$ is the Boltzmann constant, independent of the weak link $G$. The absorber linear dimensions can range from few hundred micron (micro-calorimeters) to few centimeters (macro-calorimeters) with rise time from few $\mu$s to few ms, respectively.

The performance of the sensor matters significantly to achieve a high-energy resolution with low temperature calorimetric measurements. In recent years, a lot of technologies have been developed, but three of them play a major role in neutrino experiments: Si- or Ge-thermistors, Transition edge sensors (TESs), and Magnetic Micro Calorimeters (MMCs). All of them provide a very high efficiency and energy resolution of the order of \textperthousand\ in the keV or in the MeV energy range.

In semiconductor thermistors working at cryogenics temperature, the resistivity dependence on temperature is governed by an exponential law $\rho(T) =\rho_0\,e\scriptscriptstyle^{(T_0/T)^{1/2}}$, where $T_0$ and $\rho_0$ are parameters controlled by the doping level (figure \ref{fig:sensors}, left column). Arrays of micro- and macro-calorimeters based on these devices have been widely used. Micro-calorimeters for X-ray spectroscopy achieved energy resolutions lower than 5\,eV\,@\,6\,keV with tin or HgTe absorbers~\cite{McCammon2005}, while macro-calorimeters with Neutron Transmutation Doped Ge-Thermistor for neutrinoless double $\beta$-decay achieved energy resolutions of 5\,keV at the 2615\,keV \textsuperscript{208}Tl $\gamma$ line~\cite{CUORE0Analysis}. 

TESs are superconducting thin films, operated just below its transition temperature ($T_c$). The energy deposition can drive the superconducting TES to a normal state, hence a sharp change in resistance can be sensed in TES even for a small change in temperature (figure \ref{fig:sensors}, center column). TESs can be two orders of magnitude more sensitive than the semiconductor thermistors but, since the resistance is very low (tens of m$\Omega$), they cannot be readout by conventional FET amplifier and they must be coupled with Superconducting Quantum Interference Devices (SQUID) current amplifiers. TESs are widely used in large microcalorimeter arrays for X-ray spectroscopy achieving record resolutions below 2\,eV at 1.5\,keV, with Gold and Bismuth absorber~\cite{Ullom2015}. Their high energy and time resolutions make TESs ideal detectors for developing microcalorimeters array for the direct measurements of the neutrino mass~\cite{Nucciotti_LTD}. Moreover, since TESs have a very low energy threshold, that is key requirement for light Dark Matter research, they are used in several cryogenic dark matter experiments.

MMC sensors are quite different from the previous two. They are paramagnetic sensors exposed to a small magnetic field. The temperature rise  causes a change in the sensor magnetization, which is sensed by a SQUID magnetometer (figure \ref{fig:sensors}, right column). The MMCs use the paramagnetic susceptibility of gold doped with a low concentration of erbium ions (Au:Er) when placed in a DC magnetic field. In a paramagnet, the magnetization is inversely proportional
to temperature, making it very sensitive to small changes at low temperatures. MMCs provide very high energy resolutions, i.e. 1.7\,eV\,@\,6\,keV, and they are used as micro-calorimeters for X-ray spectroscopy~\cite{Porst2014} and neutrino mass measurements~\cite{ECHo} and as macro-calorimeters for the search of the neutrinoless double $\beta$-decay~\cite{AMoRE}.

\section{Macro-calorimeter for neutrinoless double $\beta$-decay}
Double $\beta$-decay is a rare process in which a nucleus changes its atomic number by two units. It is possible only in nuclei with an even number of neutrons and protons, in which the single $\beta$-decay is energetically forbidden. In neutrinoless double $\beta$-decay only the two electrons are emitted~\cite{DellOro2016}. The decay violates the lepton number conservation by two units and it is possible only if the neutrino is a massive Majorana particle. Its transition width is proportional to the square of the effective Majorana mass $m_{\beta\beta}$.  The searches for a $0\nu\beta\beta$ signal rely on the detection of the two emitted electrons: a monochromatic line at the decay $Q_{\beta\beta}$-value. 

Since the two-neutrino double $\beta$-decay ($2\nu\beta\beta$) allowed by Standard Mode is an unavoidable source of background for $0\nu\beta\beta$ searches, detectors with excellent energy resolutions are fundamental. From simple statistical considerations, the sensitivity $S^{0\nu}$ of $0\nu\beta\beta$ decay search is given by $S^{0\nu}\propto\varepsilon\,\eta\sqrt{(M\,t)/(B\,\Delta E)}$ where $\varepsilon$, $\eta$, $M$, $t$ , $\Delta E$ and $B$ are the detector efficiency, the active isotope abundance, the source mass, the measuring time, the energy resolution and specific background at $Q_{\beta\beta}$, respectively.

There are two experimental approaches for $0\nu\beta\beta$ searches: active source external to the detector and active source internal to the detector (\textit{calorimeter}). Low temperature macro-calorimeters are particularly suitable for the second approach when the decaying isotope under study can be embedded in the absorber material. They provide high energy resolution and large flexibility in the choice of the sensitive material. Several interesting calorimetric candidates have been proposed and tested. By far the most advanced results are obtained with TeO\textsubscript{2} low-temperature macro-calorimeters. With this compound, it is possible to grow large crystals with reasonable mechanical and thermal properties together with a very large content of the candidate $0\nu\beta\beta$ \textsuperscript{130}Te (isotopic abundance: $\eta=34.167$\%). The $Q_{\beta\beta}$-value of the decay is around 2528\,keV, falling between the peak and the Compton edge of the 2615\,keV gamma line of \textsuperscript{208}Tl, the highest-energy gamma from the natural decay chains; this leaves a relatively clean window to search for the signal. 

The CUORE experiment is based on TeO\textsubscript{2} macro-calorimeters and represents the largest thermal detector array ever built~\cite{CUORE}. The detector is composed by 988 calorimeter, arranged in 19 towers of 52 crystals, for total mass of 741\,kg. Each tower consists of 13 floors of 4 crystals each. The tower structure is made of ultra pure copper and the crystals are coupled to it by mean of small PTFE supports. The absorber is a TeO\textsubscript{2} cubic crystal, $5\times 5\times × 5$\,cm\textsuperscript{3} in size, with an NTD-Ge thermistor glued onto the crystal surface. The CUORE technology was already tested during the last ten years. In particular CUORE-0 was a single CUORE-like tower built using the low-background assembly techniques developed for CUORE and running at LNGS from 2013 to 2015~\cite{CUORE0}. The average energy resolution in the Region Of Interest, defined as the FWHM of the 2615\,keV $\gamma$-ray peak, resulted 4.9\,keV, with a corresponding RMS of 2.9\,keV. This demonstrates that the CUORE goal of 5\,keV of energy resolution is feasible. The background in the ROI resulted $B = 0.058$\,cnts/keV/kg/yr. With these data, the collaboration sets a 90\% C.L. lower bound on the decay half-life of $T^{0\nu}>2.7\cdot 10^{24}$\,yr. Combining this with the exposure of \textsuperscript{130}Te from the past Cuoricino experiment, the collaboration sets a limit of $T^{0\nu} > 4.0\cdot 10^{24}$\,yr (90\% C.L.). This corresponds to a limit range on the effective Majorana mass of $m_{\beta\beta}<270-760$\,meV, using the most up-to-date NME calculations. The expected upper limit on surface-related backgrounds in the ROI in CUORE, extrapolated from test results and the performance of CUORE-0, is around $10^{2}$ counts/keV/kg/y. The projected $T^{0\nu}$ sensitivity after five years of live-time is $T^{0\nu}>9.5 \cdot 10^{25}$\,y at the 90\% C.L., which corresponds to an upper limit on the effective Majorana mass in the range $m_{\beta\beta}<50-130$\,meV, approaching the Inverted Hierarchy region~\cite{CUORE}. After successful commissioning of the CUORE cryostat, the 19 towers have been deployed in the cryostat in August of 2016. The begin operation at base temperature are currently in progress and the first data are foreseen for mid-2017.

\section{Micro-calorimeter for direct neutrino mass measurements}
Experiments based on kinematic analysis of electrons emitted in single $\beta$-decay, are the only ones dedicated to effective electron-neutrino mass determination.
The method consists in searching for a tiny deformation caused by a non-zero neutrino mass to the spectrum near its end point. The most stringent result is $m_{\beta}<2\,\mbox{eV}$\,@\,90\%\,C.L. and comes from electrostatic spectrometers on tritium decay ($Q=18.6$~keV)~\cite{Troitsk2001,Mainz2004}. KATRIN, a next generation experiment, is designed to reach a sensitivity of $m_\beta<0.2~eV$ in five years~\cite{KATRIN}. With KATRIN, this experimental approach reaches its technical limits and it is therefore mandatory to develop alternative and complementary experimental methodologies for the direct neutrino mass measurements. The micro-calorimetric measurement, where the $\beta$-source is embedded in the detector absorber, is a promising alternative. 

Two experiments investigated the feasibility of micro-calorimeter based on the \textsuperscript{187}Re isotope: MANU~\cite{MANU} and MIBETA~\cite{MIBETA}. The first used one detector with NTD-Ge thermistor glued on a 1.6\,mg single crystal of metallic rhenium while the second used an array of eight Si:P implanted Si-thermistors with AgReO\textsubscript{4} for a total mass of 2.2\,mg. Both collected a statistic of around $10^7$ $\beta$-events, corresponding to a neutrino mass sensitivity of $m_\beta<20$\,eV. To improve this limits by a factor 100, so that to reach the sub-eV sensitivity needed to approach the Inverted Hierarchy, a statistic increase of $10^8$ is required. One of the most interesting candidate as an alternative to \textsuperscript{187}Re is \textsuperscript{163}Ho. The calorimetric measurement of the \textsuperscript{163}Ho was suggested in 1982 by De Rujula and Lusignoli~\cite{DeRujula} but only in recent times low temperature microcalorimeters reached the necessary maturity to be used in a large scale experiment with the potential to reach the sub-eV neutrino mass sensitivity. In a $^{163}$Ho calorimetric experiment the statistical sensitivity on the neutrino mass is proportional to $\sqrt[4]{1/N_{ev}}$, where $N_{ev}$ is the number of the detected events. Considering micro-calorimeters with $^{163}$Ho-embedded absorber, with high energy and time resolutions ($\simeq\,1$\,eV and $\simeq\,1$\,$\mu$s, respectively) and a pile-up fraction within the range $f_{pp}=10^{-3}-10^{-6}$, a sub-eV neutrino sensitivity can be reached collecting a total number of events around $N_{ev}> 10^{17}-10^{19}$~\cite{Nucciotti_Sensitivity}. To achieve these requirements a Megapixel arrays experiment is needed, where the multiplexing factor (number of multiplexed detector signals) and the read-out bandwidth play a crucial role for future developments. Currently, three experiments explore the approach of using electron capture on \textsuperscript{163}Ho to assess the neutrino mass: ECHo~\cite{ECHo}, HOLMES~\cite{HOLMES}, and NuMECS~\cite{NuMECS1}.

ECHo employs MMCs sensor with microwave multiplexing. The holmium source is enclosed in gold absorbers. With their first prototype, ECHo could demonstrate excellent energy resolution of 7.6\,eV\,@\,6\,keV and fast rise times of $\tau_r = 130\,$\,ns. A larger detector array with 16 pixels, increased purity and activity (0.1\,Bq) is being tested at the moment. In the next step ECHo-1k will exploit 100 pixels to collect $10^{10}$ events in 1 year, pushing the neutrino mass sensitivity of below 10\,eV. In the final step ECHo-1M will be composed by $10^4$ pixels with the aim to reach a sensitivity below 1\,eV in 1 year. HOLMES is making use of the TES technology with microwave multiplexing. The collaboration is currently performing the detector optimization and the read-out R\&D development. A custom ion-implanter is being assembled to embed the \textsuperscript{163}Ho in the detectors. First result on the developed detectors showed an energy resolution of about 6\,eV\,@\,6\,keV and a rise time of $\tau_r \simeq 10\,\mu$s, fulfilling the HOLMES requirement~\cite{HOLMES}. The first tests with first implanted sub-arrays will start in 2017, while the commissioning of the final 1024 pixels array will be performed in 2018. The final goal is to reach a neutrino mass statistical sensitivity below 2\,eV. NuMECS is also pursuing the TES technology. The collaboration focus is on \textsuperscript{163}Ho production via proton activation of dysprosium, as opposed to the most common neutron irradiation on \textsuperscript{162}Er. In their prototype the source nuclei are enclosed as liquid drop in a nanoporous gold absorber. NuMECS successfully measured a \textsuperscript{163}Ho spectrum with an energy resolution of about 40\,eV FWHM. NuMECS future plans include the deployment of four 1024 pixel arrays, aiming at a statistical sensitivity of about 1\,eV.

\section{Conclusion}
The study of frontier subjects in neutrino physics requires the use of frontier detectors. Experiments with large number of pixel, large mass and very high efficiency and energy resolutions, of the order of \textperthousand\ in the MeV or in the keV energy range, are a fundamental tool to study and understand the neutrino properties. 
The goal of the next future experiments is to approach and cover the Inverted Hierarchy region.
Low-temperature micro and macro calorimeters today are mature devices which often represent the only technical opportunity to satisfy these challenging demands.

\section*{References}
\bibliography{giachero.7YRM.proceedings}

\end{document}